\documentclass{article}

\usepackage{microtype}
\usepackage{graphicx}
\usepackage{subfigure}
\usepackage{booktabs} 
\usepackage{multirow}
\usepackage[acronym,nonumberlist]{glossaries}
 \newglossaryentry{Fault}
{
    name=Fault,
    description={Condition of a machine that occurs when one of its components or assemblies degrades or exhibits abnormal behavior}
}

 \newglossaryentry{Failure}
{
    name=Failure, 
    description={Termination of the ability of a component to perform a required function. Normally a failure event is preceded by a fault state}
}

\newacronym{om}{O\&M}{Operation and Maintenance}

\newacronym{scada}{SCADA}{Supervisory Control and Data Acquisition}

\newacronym{cm}{CM}{Condition Monitoring}

\newacronym{cms}{CMS}{Condition Monitoring Systems}

\newacronym{nbm}{NBM}{Normal Behaviour Model}

\newacronym{anbm}{ANBM}{Autoregressive Normal Behaviour Model}

\newacronym{pdm}{PdM}{Predictive Maintenance}

\newacronym{fp}{FP}{False Positive}

\newacronym{tp}{TP}{True Positive}

\newacronym{fn}{FN}{False Negative}

\newacronym{tn}{TN}{True Negative}

\newacronym{cnbm}{CNBM}{Causal Normal Behaviour Model}

\newacronym{snbm}{SNBM}{Simultaneous Normal Behaviour Model}

\newacronym{acnbm}{ACNBM}{Autoregressive Causal Normal Behaviour Model}

\newacronym{asnbm}{ASNBM}{Autoregressive Simultaneous Normal Behaviour Model}

\newacronym{wt}{WT}{Wind Turbine}

\newacronym{lcoe}{LCOE}{Levelized Cost of Energy}

\newacronym{hss}{HSS}{High Speed Shaft}

\newacronym{ims}{IMS}{Intermediate Speed Shaft}

\newacronym{lss}{LSS}{Low Speed Shaft}

\newacronym{gbm}{GBM}{Gradient Boosting Machine}

\newacronym{gbdt}{GBDT}{Gradient Boosting Decision Trees}

\newacronym{svm}{SVM}{Support Vector Machines}

\newacronym{ann}{ANN}{Artificial Neural Network}

\newacronym{knn}{KNN}{K-Nearest Neighbors}

\newacronym{gps}{GPs}{Gaussian Processes} 

\usepackage{hyperref}

\usepackage[accepted]{icml2019}

\icmltitlerunning{The Impact of Feature Causality  on Normal Behaviour Models for SCADA-based Wind Turbine Fault Detection}

\begin{document}

\twocolumn[
\icmltitle{The Impact of Feature Causality  on Normal Behaviour Models for SCADA-based Wind Turbine Fault Detection}



\icmlsetsymbol{equal}{*}

\begin{icmlauthorlist}
\icmlauthor{Telmo Felgueira}{ist,jungle}
\icmlauthor{Silvio Rodrigues}{jungle}
\icmlauthor{Christian S. Perone}{jungle}
\icmlauthor{Rui Castro}{inesc}
\end{icmlauthorlist}

\icmlaffiliation{ist}{Department of Electrical and Computer Engineering, Instituto Superior Tecnico, Lisbon, Portugal}
\icmlaffiliation{jungle}{Jungle.ai, Lisbon, Portugal}
\icmlaffiliation{inesc}{INESC-ID/IST, University of Lisbon, Portugal}

\icmlcorrespondingauthor{Telmo Felgueira}{telmo.felgueira@jungle.ai}

\icmlkeywords{Machine Learning, ICML}

\vskip 0.3in
]




\begin{abstract}
The cost of wind energy can be reduced by using SCADA data to detect faults in wind turbine components. Normal behavior models are one of the main fault detection approaches, but there is a lack of consensus in how different input features affect the results. In this work, a new taxonomy based on the causal relations between the input features and the target is presented. Based on this taxonomy, the impact of different input feature configurations on the modelling and fault detection performance is evaluated. To this end, a framework that formulates the detection of faults as a classification problem is also presented. 

\end{abstract}

\section{Introduction}
\label{submission}

In 2018, global energy-related CO$_2$ emissions reached a historic high of 33.1 gigatonnes. These emissions are caused by the burning of fossil fuels, mainly natural gas, coal and oil, which accounted for 64\% of global electricity production in this same year \cite{iea:co2}. Greenhouse gases like CO$_2$ are responsible for climate change which threatens to change the way we have come to know Earth and human life. For the previous reasons, there has been a global effort to shift from a fossil fuel based energy system towards a renewable energy one. In fact, it is expected that by 2050 wind energy will represent 14\% of the world's total primary energy supply \cite{dnv:data}.

The operation and maintenance costs of \glspl{wt} can account for up to 30\% of the cost of wind energy \cite{ewea:om}. This happens because while generators in fossil fuel power plants operate in a constant, narrow range of speeds, \glspl{wt} are designed to operate under a wide range of wind speeds and weather conditions. This means that stresses on components are significantly higher, which increases the number of failures and consequently the maintenance costs. 

There have been recent efforts to monitor and detect incipient faults in \glspl{wt} by harvesting the high amounts of data already generated by their \gls{scada} systems, which, in turn, enables the wind farm owners to employ a predictive maintenance strategy. In fact, it is expected that by 2025 new predictive maintenance strategies can reduce the cost of wind energy by as much as 25\% \cite{irena:change}. One of the main methods for monitoring the condition of \glspl{wt} is building \glspl{nbm} of the component temperatures. The fundamental assumption behind the use of \glspl{nbm} is that a fault condition is normally characterized by a loss of efficiency, which results in increased temperatures.  By using SCADA data to build a model of the temperatures of the \gls{wt} components, one can calculate the residuals, which are the difference between the real values measured by the sensors and the predicted values by the model. These residuals can be used to detect abnormally high temperatures that may be indicative of an incipient fault. 

Multiple works \citep{article:zaher, article:brandao1, inproceedings:brandao2} have reported good results using \glspl{nbm} to predict \gls{wt} failures, being able to predict failures in \gls{wt} components months in advance. In these works the authors used as features active power, nacelle temperature and lagged values of the target temperature, thus including autoregressive properties into to the model, to predict the temperatures of various components. In \cite{meik:0} and \cite{article:bach} the authors obtained an important result: although the use of autoregressive features resulted in better temperature modelling performance it also resulted in worse fault detection performance. Another important result was obtained in \cite{bang:1} and \cite{tautz:phd}, which indicated that using features that are highly correlated with the target also increased the modelling performance but decreased the fault detection performance of the model. Nonetheless, this type of features are still used in many works today, such as \citep{article:bach, bach:phd, today:colone, today:zhao, today:tautz, today:zhao2, today:mazdi}. There are also conflicting opinions regarding the use of autoregressive features, with some works using them and others not. The main reason behind this is the lack of consistent case studies that evaluate the impact of different features on both the temperature modelling and fault detection performances. It should also be noted that in \glspl{nbm} it's not trivial that the more features the model has the better its fault detection performance will be. This happens because the model is being trained to minimize the temperature modelling error and not the fault detection one. Having this in mind, this work will present a new feature taxonomy to distinguish different input feature types. Then, the impact of these input feature types on the temperature modelling and fault detection performances will be evaluated.


Finally, evaluating the fault detection performance of different models is not as trivial as evaluating their temperature modelling performance. In fact, there is no standard in the literature regarding how to evaluate fault detection performance. This happens because of the inherent nature of the fault detection problem, in which there is rarely groundtruth. Indeed, there is data of when the failure happened, but there is no information regarding when the fault state started, making it not trivial to formulate as a classification problem. Hence why the majority of the literature evaluates the fault detection results by visual inspection, observing the increase in the residuals before the failure. This is problematic, because comparisons between different models will be highly subjective. Having this in mind, this work will also present a formulation of the detection of faults as a classification problem.

\section{Methods}

\subsection{Data and Training }

In this work a dataset composed of 15 turbines during a 6 year period will be used. This data corresponds to SCADA signals with 10 minute resolution. During the year of 2012 there was a total of 5 failures related with the Gearbox IMS Bearing. For these reasons, this will be the component for which an \gls{nbm} will be trained,  with the objective of predicting the corresponding failures. 

The models will be trained with data from the beginning of 2007 to the end of 2011 and tested on data from 2012. Periods with faults will be removed from the training data so the model does not learn abnormal behaviour. The models will be implemented with \gls{gbdt}, which work by iteratively combining weak decision trees into a strong ensemble learner. In terms of implementation,  LightGBM \cite{lightgbm} will be used due to its high computational performance. In terms of optimization, the year of 2011 will be used as a validation set when choosing the number of trees for each model by early stopping. Note that no exhaustive hyperparameter optimization was performed, so all models will use the same hyperparameters besides the number of trees.


\subsection{Feature Taxonomy}

In the present work we hypothesize that what causes a decrease in fault detection performance is not using input features highly correlated with the target, but using those whose sensors are physically close to the target sensor. If there is an increase in the temperature of a faulty component, the physically close components will also get hotter due to heat transfer. Thus, using physically close components as features to the model may leak information regarding the fault state of the target, making it unable to detect abnormal behaviour. These ideas can be clarified by using appropriate nomenclature. Based on Econometric Causality \cite{econ:nom}, we will distinguish features based on their causal relations with the target. If the target is causally dependent of the features, they are causal features. On the other hand, if the target depends on the features but the features also depend on the target they are simultaneity features. Such causal relations are assumed based on the domain knowledge of the physical system. 

Based on the taxonomy previously presented, different models will be defined based on their input feature configuration. The simplest model that will be tested is the \gls{cnbm}, which only uses causal features. These are determined based on domain knowledge and will be: rotor speed, active power, pitch angle, wind speed and ambient temperature. All these features characterize the operation regimes of the \gls{wt}, these are causal features because the gearbox IMS bearing temperature depends on their values, but their values are not dependent on it. For example, variations in the ambient temperature influence the gearbox IMS bearing temperature, but the influences of the latter on the ambient temperature can be disregarded.

On the other hand, simultaneity features will be chosen based on Pearson Correlation, which is a standard first approach for regression problems. The highest correlated feature with the gearbox IMS bearing temperature is the gearbox HSS bearing temperature, which is a simultaneity feature because there is heat transfer between the two sensors, thus meaning that their values are mutually causally dependent. Having this in mind, the \gls{snbm} will use all the features from the \gls{cnbm} plus gearbox HSS bearing temperature. Two more models will be tested, which correspond to the autoregressive versions of the previously described models: \gls{acnbm} and \gls{asnbm}. 



\subsection{Fault Evaluation Framework}

\begin{figure}[t]
  \centering
  \includegraphics[trim= 0 0 0 5 ,clip,width=0.5\textwidth]{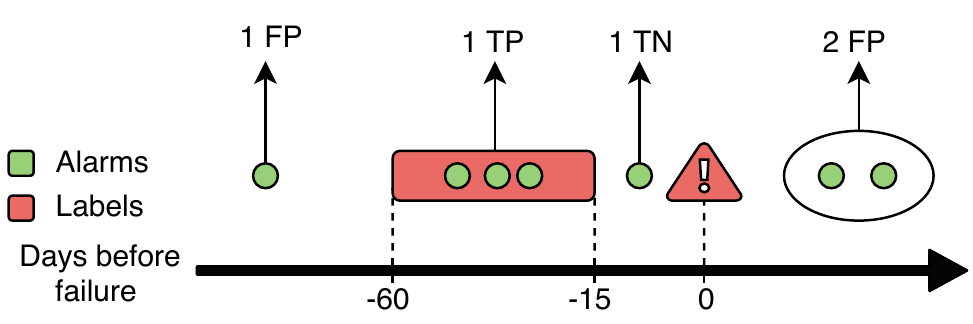}
   \caption{Schematic example of fault detection formulated as a classification problem. } 
   \label{fig:evalfram}
\end{figure}

To develop an evaluation framework for fault detection, one must first formulate it as a binary classification problem where there are two labels: fault and no-fault. Since there is no information regarding the fault state of the component, only the date of failure, it was defined with the wind farm owners that for the failures studied in this work it can be assumed that a fault state would be present at most 60 days before the failure. It was also defined that for the alarms to be useful they should be triggered at least 15 days before the failure. This means that to be considered a \gls{tp} the alarm must be triggered between 60 and 15 days before the failure. Figure \ref{fig:evalfram} presents a schematic example of the previously described problem formulation. Taking this example, it is important to note that the number of alarms triggered in the prediction window is not relevant, they are all aggregated as 1 \gls{tp}. The main reason for this, is that if the aggregation is not done, then 4 alarms for the same failure would count as much as 4 detected failures with 1 alarm each. This clearly is not what is intended of the framework, since 1 alarm should be enough to motivate an inspection, and detecting 4 failures with 1 alarm outweighs
detecting 1 failure with 4 alarms. Finally, it is also important to note that alarms triggered less than 15 days before the failure are not considered \glspl{fp}, since there is indeed a fault state, it simply is not relevant, so they are considered \glspl{tn}.


\section{Results}

In terms of temperature modelling, the models were evaluated on periods of turbines that are known to be healthy. The results, presented in Table \ref{resulstnbm}, indicate that the use of simultaneity features indeed improves the modelling performance, since SNBM obtains better results than CNBM. The use of autoregressive features also improves the modelling performance, since ACNBM and ASNBM obtain better results than their non-autoregressive counterparts. This results make sense, since there are certain regimes of the turbine that are difficult to model without simultaneity nor autoregressive features, such as the turning off of the turbine as noted in \cite{bach:phd}.

\begin{table}[t]
\caption{Regression error metrics for the training and test sets of each model.}
\label{resulstnbm}
\vskip 0.15in
\begin{center}
\begin{small}
\begin{sc}
\begin{tabular}{lcccr}
\toprule
\multirow{2}{*}{Model} & \multicolumn{2}{c}{Training} & \multicolumn{2}{c}{Test} \\
                       & MAE          & RMSE          & MAE         & RMSE       \\
\midrule
CNBM                   &        1.48      &     2.14          & 1.80        & 2.62       \\
SNBM                   &    0.87          & 1.26              & 1.01        & 1.41       \\
ACNBM                  &    1.03          & 1.57              & 1.14        & 1.67       \\
ASNBM                 & 0.83             & 1.22              & 0.96        & 1.38      \\
\bottomrule
\end{tabular}
\end{sc}
\end{small}
\end{center}
\vskip -0.1in
\end{table}
 
\begin{figure}[t]
  \centering
  \includegraphics[trim= 5 15 40 70 ,clip,width=0.5\textwidth]{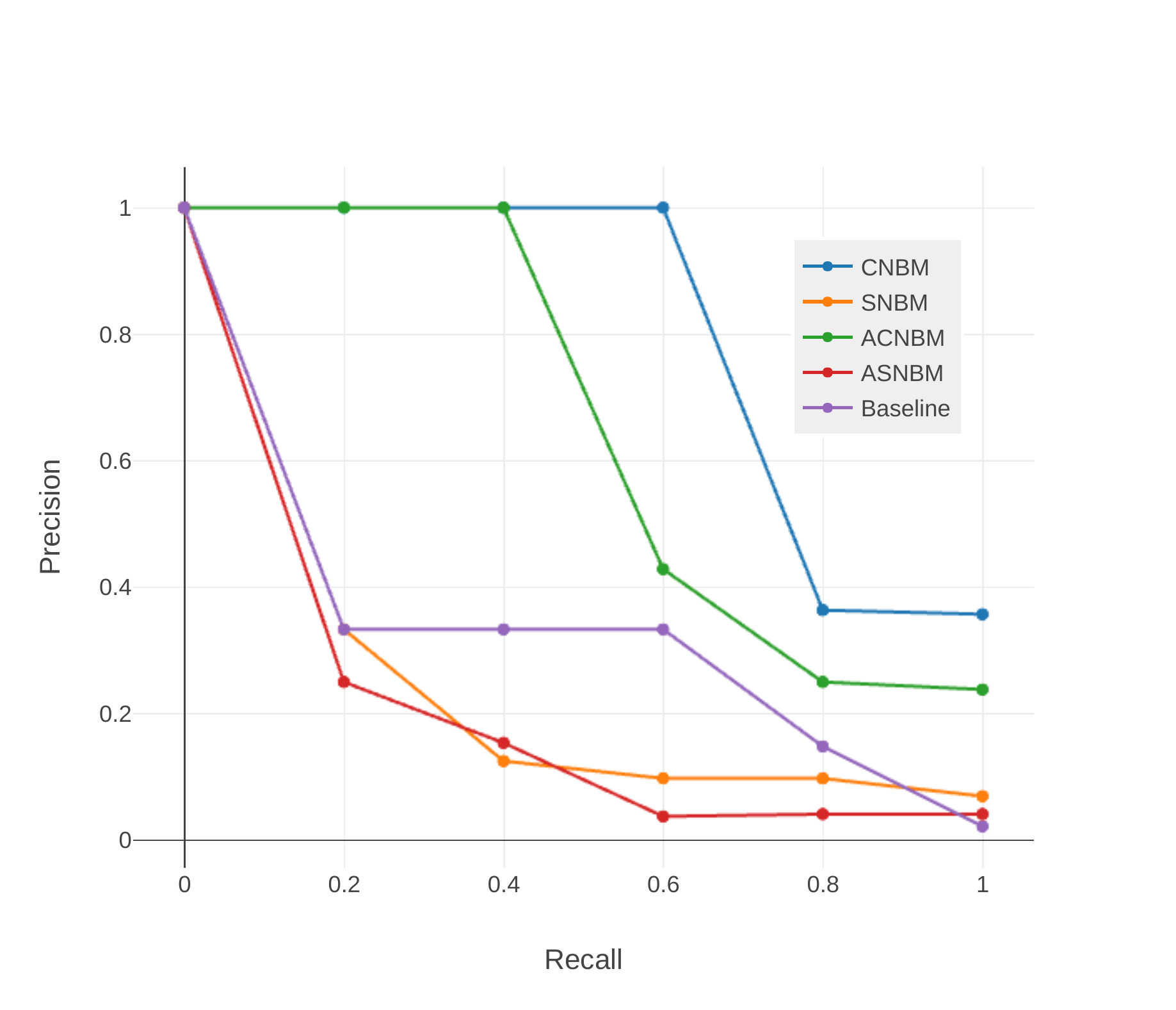}
   \caption{Precision and Recall curves for the different models.} 
   \label{fig:results}
\end{figure}

In terms of fault detection, a baseline was defined that consists of setting different thresholds on the distribution of the target temperature and obtaining the corresponding precision and recall. For the models, also different thresholds were applied in the residuals to obtain the different values of precision and recall. The results are presented in Figure \ref{fig:results}. As can be seen, the CNBM which obtained the worst modelling performance obtains the best fault detection performance. Also, note that the models with the simultaneity feature are significantly worse than the baseline. 

\section{Conclusions}

An evaluation framework to formulate fault detection as a classification problem was presented. This hopes to contribute to the development of a standard approach for fault detection performance evaluation. Furthermore, a taxonomy regarding the causal relations of the different input feature types was presented, which hopes to make the discussion on how different features affect the performance of models clearer. Finally, it was demonstrated that although autoregressive and simultaneity features increase the modelling performance they decrease the fault detection capabilities of the model. This is an important contribution since the majority of works today still use these types of features.

%
%


\bibliography{biblio}
\bibliographystyle{icml2019}

%
%
%

\end{document}